\documentclass[onecolumn,traditabstract]{aa}
\usepackage{natbib}
\usepackage{graphicx}
\bibpunct{(}{)}{;}{a}{}{,}
\newcommand{\beq}{\begin{equation}}
\newcommand{\eeq}{\end{equation}}
\begin{document}
\title{Phase Shift Sequences for an Adding Interferometer}
\author{Peter Hyland\inst{1}\thanks{Current address: Physics Department,
McGill University, Montr\'eal, QC, Canada H3A 2T8}
\and Brent Follin\inst{2} \and Emory F. Bunn\inst{2}}
\institute{Physics Department, University of
Wisconsin - Madison, Madison, WI 53706 \and Physics Department,
University of Richmond, Richmond, VA 23173}
\titlerunning{Phase shift sequences}
\authorrunning{Hyland et al.}

\abstract{Cosmic microwave background (CMB) polarimetry has the
potential to provide revolutionary advances in cosmology.  Future
experiments to detect the very weak B mode signal in CMB
polarization maps will require unprecedented sensitivity and control
of systematic errors.  Bolometric interferometry may provide a way
to achieve these goals. In a bolometric interferometer (or other
adding interferometer), phase shift sequences are applied to the
inputs in order to recover the visibilities.  Noise is minimized
when the phase shift sequences corresponding to all visibilities are
orthogonal.  We present a systematic method for finding sequences
that produce this orthogonality, approximately minimizing both the
length of the time sequence and the number of discrete phase shift
values required. When some baselines are geometrically equivalent,
we can choose sequences that read out those baselines
simultaneously, which has been shown to improve signal to noise
ratio. }


\date{\today}
\maketitle

\section{Introduction}
The field of observational cosmology has been advancing quickly in
recent years. Observations of the cosmic microwave background (CMB)
radiation have been leading the way, as evidenced by WMAP's highly
successful mapping of CMB anisotropy \citep{wmap5yrbasic}, DASI's
detection of the polarized component of the CMB \citep{dasi}, and
the 2006 Nobel Prize in Physics awarded to John Mather and George
Smoot. Momentum is building for experiments that characterize the
CMB polarization in detail \citep{weissreport}.

A linear polarization map can always be expressed as the sum of two
component maps, denoted E and B \citep{selzal,kks}. CMB experiments
to date have detected only the ``curl-free'' E component, which is
produced primarily by (scalar) density perturbations. The
``divergence-free'' B component is not produced by scalar
perturbations at linear order, and is therefore a clean probe of
other, smaller effects. In particular, inflationary models predict a
B-mode signal produced by gravitational wave (tensor) perturbations
in the early universe. These B modes promise to hold key information
about the process of inflation and particle physics above the Grand
Unification scale. The challenge of finding the B modes is no small
task, however: the B component is expected to be at least an order
of magnitude weaker than the E component (which is itself small
compared to the temperature anisotropy) over all angular scales.
Experiments to search for B modes will require unprecedented
sensitivity and control of systematic errors.

Bolometric interferometry is one proposed method for achieving these
goals \citep{brain1,brain2,mbi1,mbi2,tucker:70201M}. A bolometric
interferometer is a marriage between highly sensitive, incoherent
bolometric detectors and the phase-sensitive,
systematic-error-reducing observing technique of interferometry.
\citet{hamilton} have shown that a bolometric interferometer can
achieve sensitivities comparable to traditional technologies.  The
question of whether bolometric interferometry is useful for CMB
polarimetry will thus depend on the method's ability to control
systematic errors. Systematic errors in interferometers are
certainly different from those in imaging experiments \citep{bunn};
it can be argued that interferometers are superior in this regard,
although this question requires further research.

In a bolometric interferometer, the signals from a set of input
feedhorns are combined with either a Butler combiner or a
quasi-optical (Fizeau) combiner.  In either case, bolometers measure
the total power in the combined beam -- that is, each bolometer is
illuminated by signals from all of the input horns.  Since the
signal in each detector is the sum of all the inputs, a bolometric
interferometer is an example of an ``adding'' interferometer (as
opposed to traditional radio interferometers, which are
``multiplying'' interferometers). One of the keys to making this
method work is to arrange for the phase information to be encoded in
the bolometer signals, so that individual pairwise visibilities can
be extracted. To achieve this goal, a sequence of phase shifts can
be applied to each of the input horns, in such a way that each
visibility is phase-shifted in an independent fashion.  The
resulting time series can be solved for the individual visibilities.
The phase shift sequences should be chosen so that this inversion
can be done with minimal noise.

We would like the length of the phase shift sequence to be as short as
possible, to avoid error due to $1/f$ noise in the detectors.
Clearly the number of
phase shifts must be at least as large as the number of visibilities
to be recovered.  In the most general case, an $N$-horn interferometer
has $N(N-1)/2$ distinct visibilities, requiring long phase shift sequences
for interferometers with many inputs.  On the other hand, if the
input horns are arranged in a regular pattern, such as a square array,
then many antenna pairs correspond to identical visibilities.  These
can be given identical phase shifts and read out together.  This
coherent treatment of equivalent (or redundant) baselines has
two advantages. First, it allows for shorter phase shift
sequences.  Second, by coadding equivalent signals, the signal-to-noise
ratio
is improved \citep[hereinafter C08]{charlassier}.

C08 gave an excellent overview of
how a bolometric interferometer works and considered the choice of
phase shift sequences in detail.  For the case of a square array of
horns, the paper presented a method of phase modulating the inputs
that gives equivalent baselines identical phase shift sequences. In
this paper we present independently-developed work on phase modulation
and coherent addition of
baselines that complements
the methods of C08. We consider general horn arrangements as well as a
regular square lattice.  In the general case, we present a method
for finding the optimal phase shift sequence assuming no baselines
are redundant.  In the case of a square array, we present a
refinement of the method of C08.  Unlike the original method, which
achieves optimal noise performance only in the limit as the number
of time steps tends to infinity, our method is optimal for
sequences of nearly or exactly the minimum possible length.

In Sect.~\ref{sec:Form} we present our formalism for denoting
sequences of phase shifts and consider the criterion for an optimal
phase shift sequence.  Section \ref{sec:Meth} introduces a
shorthand notation and applies this to a method for constructing
bases for phase sequences and selecting optimal sequences for
interferometers without equivalent baselines. In
Sect.~\ref{sec:Array} we consider a square array of horns,
accounting for redundant baselines. For simplicity, we consider only
one linear polarization state in these sections, but in
Sect.~\ref{sec:Pol} we generalize to an array with two
polarizations.  Section \ref{sec:Conclusions} contains a brief
concluding discussion, and Appendix A contains a
useful mathematical result.

\section{Formalism}
\label{sec:Form} Suppose that our interferometer has $N$ input
horns, each of which receives one electric field component.  (We
will consider the case where both $x$ and $y$ components are
received in Sect.~\ref{sec:Pol}.)  Assuming monochromatic radiation
of frequency $\nu$ for simplicity, the signal entering the $j$th
horn can be written $E_je^{2\pi i\nu t}$. This signal is presumed
to have already been averaged over the antenna pattern, so no integral
over the sky position will be explicitly shown in the equations below.
We apply a time-dependent
phase shift $\phi_j(t)$ to each of these inputs. Since bolometers
measure the total power, the signal detected by the $m$th detector
is
\beq S_m(t)\propto\left|\sum_{j=1}^N E_j
e^{i(\Delta_{jm}+\phi_j(t))}\right|^2 =\sum_{j,k=1}^n E_jE_k^*
e^{i(\Delta_{jm}-\Delta_{km}+ \phi_j(t)-\phi_k(t))}, \label{eq:Soft}
\eeq
where the phase shifts $\Delta_{jm}$ are fixed by the geometry
of the system and do not vary in time. In this expression, the
detector can correspond either to a single point on the focal plane
of a quasioptical combiner or to a single output of a Butler
combiner.

We are assuming here that all inputs contribute to all
detectors with equal amplitude.  If this assumption is relaxed, then
an additional real factor $A_{jm}$ would need to be included in each term of
the sum.
The presence of these factors would affect the overall sensitivity
of the detector to the various visibilities, but we do not expect
it to influence the optimal choice of phase shifts, so we omit it.

Let us assume for the moment that we wish to recover the visibility
associated with each pair of horns separately; we will return below
to the case in which redundant baselines are coherently added before
detection. We wish to choose the phase shifts $\phi_j(t)$ to enable
recovery of all of the cross terms $E_jE_k^*$ from each detector. In
principle, we could aim for a weaker goal, namely to ensure that
each cross term be recoverable from the full set of detector outputs
$S_1(t),S_2(t),\ldots$; however, to avoid systematic errors
resulting from subtracting signals in different detectors, it is
preferable to insist that each visibility be recovered from each
detector separately.

Since we are focusing on one detector at a time, we suppress the
subscript $m$ in eq.~(\ref{eq:Soft}).  Furthermore, the
time-independent phase shifts $\Delta_{jm}$ do not affect the
problem of visibility recovery, so we suppress these as well.
Finally, we assume that the phase shifts $\phi_j$ are changed in
discrete time steps, so we replace the functions $\phi_j(t)$ with
sequences $\phi_{jt}$.  Here $t=1,2,\ldots,M$, where $M$ is the
number of steps in the phase shift sequence at each detector. With
these changes, equation (\ref{eq:Soft}) becomes
\beq
S_t\propto
\sum_{j,k=1}^NE_jE_k^* e^{i(\phi_{jt}-\phi_{kt})}\equiv
V_{jk}e^{i(\phi_{jt}-\phi_{kt})}.
\eeq
Finally, we assume that the
phase shifts can take on $P$ equally-spaced values from 0 to $2\pi$:
\beq \phi_{jt}\in \{2\pi p/P \ \ | \  \ p=0,1,2,\ldots,P-1\}.
\eeq

Given the time sequence of measurements $S_1,S_2,\ldots,S_M$, we
wish to recover all $N(N-1)$
complex visibilities $V_{jk}\equiv E_jE_k^*$
with $j\ne k$
(or equivalently to recover both real and imaginary
parts of all pairs with $j<k$).
In addition, we will always recover
the total power term $ \sum_j V_{jj}=\sum_j|E_j|^2$, which enters each
$S_t$ equally. Solving for the cross terms is therefore simply
inverting a linear system of $M$ equations for $N(N-1)+1$ unknowns.
Generically, we expect this to be possible as long as
\beq
M\ge
N(N-1)+1\equiv N_{\rm vis}.
\eeq

We want to insist not just that the visibilities be recovered, but
that they be recovered with minimal possible noise. To be specific,
the recovery problem we wish to solve is
\beq
\vec S={\cal A}\vec V,
\eeq
where $\vec S$ is the $M$-dimensional signal vector, $\vec V$ is the
$N_{\rm vis}$-dimensional vector of visibilities to be recovered,
and ${\cal A}$ is a matrix whose elements are determined by the phase
shifts:
\beq
A_{tm}=e^{i(\phi_{jt}-\phi_{kt})},
\eeq where the $m$th
visibility $V_m$ corresponds to the horn pair $jk$. In this
situation, where all elements of the matrix ${\cal A}$ have absolute value
equal to one, the minimum possible noise contributions to the
visibilities is achieved when all columns of $A$ are orthogonal:
\beq
\sum_t A_{tm}^*A_{tm'}=0\qquad\mbox{ for $m\ne m'$}
\eeq
or
equivalently for ${\cal A}^\dag {\cal A}$ proportional to the identity matrix.

A proof of this statement is provided in Appendix \ref{sec:appendix}.
Intuitively,
it says that the visibilities are recovered with minimum noise when
they are maximally independent of each other, that is, when they
contribute orthogonally to the time series of signals at the
detector.

Let us summarize.  For any pair of horns $j,k$, define an
$M$-dimensional vector \beq
\vec\Phi_{jk}=(e^{i(\phi_{j1}-\phi_{k1})},e^{i(\phi_{j2}-\phi_{k2})},
\ldots,e^{i(\phi_{jM}-\phi_{kM})}). \label{eq:Phi} \eeq Our goal is
to choose the set of phase shifts $\phi_{jt}$ such that the vectors
$\vec \Phi_{jk}$ and $\vec \Phi_{j'k'}$ are orthogonal whenever
$(jk)\ne(j'k')$. When this condition is satisfied, each visibility
is recovered simply by taking the dot product of the detector signal
with the corresponding vector $\vec\Phi$: the estimator of $V_{jk}$
is
\beq \hat V_{jk}={1\over M}\vec\Phi_{jk}^\dag\vec S= {1\over
M}\sum_{t=1}^M e^{-i(\phi_{jt}- \phi_{kt})}S_t \eeq We will call the
vector $\vec\Phi_{jk}$ the ``mask'' for the baseline $jk$.

Note that $\vec\Phi_{jk}$ and $\vec\Phi_{kj}$ are complex conjugates
of each other.  The requirement that these be orthogonal, which
means roughly that the elements of $\Phi_{jk}$ uniformly sample
directions in the complex plane, is necessary for both the real and
imaginary parts of $V_{jk}$ to be recovered with minimum noise.

It may be instructive to compare the phase shift schemes for the
bolometric interferometer with those applied in  a traditional
multiplying interferometer.  In traditional interferometry,
orthogonal patterns of square-wave phase shifts (e.g., Walsh
functions) are applied to each of the input antennas in order to
reduce the response of the instrument to  spurious signals
\citep[e.g.,][]{thompson}.  The phase shift patterns we require in
the adding interferometer must obey a more stringent orthogonality
requirement: rather than merely demanding orthogonality of all of
the input phase shifts (i.e., demanding that the $\vec\phi_j$ be
orthogonal), we require that the phase shifts associated with all
{\it visibilities} (i.e., all $\vec\Phi_{jk}$) be orthogonal.

\section{Method for finding phase shifts}
\label{sec:Meth}
Let us suppose that the number $N$ of horns is
fixed, as is the number $P$ of possible phase shift values.  We wish
to find the shortest sequence of time steps (that is, the minimum
$M$) that satisfies our orthogonality criterion.  Alternatively,
given $M,P$, we can ask for the maximum number of horns that can be
accommodated.

We will introduce the following shorthand notation for the possible
phase factors: \beq [p]=e^{i 2\pi p/P},\qquad p=0,1,2,\ldots P-1.
\eeq For purposes of illustration, we will consider the case $P=4$
in this section, so that the four possible phase shift values are
\beq [0123]=(1,i,-1,-i). \eeq The method we outline generalizes to
other values of $P$.

Let the number of time steps $M$ be a power of 4: $M=4^\mu$ for some
positive integer $\mu$. We can define a set of $\mu$ mutually
orthogonal $M$-dimensional vectors as follows: the vector
$\vec\alpha_1$ is obtained by stepping through the four possible
phase values as slowly as possible -- that is, it consists of $M/4$
repetitions of [0], followed by $M/4$ repetitions of [1], etc. Each
subsequent vector $\vec\alpha_j$ cycles through the possible phases
four times faster until the last one $\vec\alpha_\mu$, which consists
of $M/4$ repetitions of the sequence $[0123]$. To be explicit, here
is the case $\mu=3$:
\beq
\vec\alpha_1=
[0000000000000000111111111111111122222222222222223333333333333333]
\eeq\beq
\vec\alpha_2=
[0000111122223333000011112222333300001111222233330000111122223333]
\eeq\beq
\vec\alpha_3=
[0123012301230123012301230123012301230123012301230123012301230123]
\eeq
Here is an alternative description of the construction
of these vectors: the $k$th element of the vector $\vec\alpha_j$
is the $j$th-most-significant digit in the base-4 expression for $(k-1)$.
In the general case with $M = P^{\mu}$, $\vec\alpha_{1}$ steps
through the $P$ values as slowly as possible and each subsequent
$\vec\alpha_{j}$ cycles $P$ times faster.

We now define
\beq
\langle j_\mu,\ldots,j_2,j_1\rangle=\vec\alpha_\mu^{j_\mu}\cdots
\vec\alpha_2^{j_2}\vec\alpha_1^{j_1}
\eeq
for integers
$j_\mu,\ldots j_2,j_1$ between 0 and 3.  Here multiplication and
exponentiation are performed elementwise in each vector.  Since
$[p]$ is shorthand for $e^{ip\pi/2}$, multiplication corresponds to
addition modulo 4 on the values in square brackets. For instance,
in the case $\mu=2$,
\begin{eqnarray}
\langle 2,1\rangle=\vec\alpha_2^2\vec\alpha_1
&=&[0123012301230123]^2[0000111122223333]\nonumber\\
&=&
[0202020202020202][0000111122223333]
=[0202131320203131].
\end{eqnarray}

It is straightforward to check that the vectors $\langle
j_\mu,\ldots,j_2,j_1\rangle$ are all mutually orthogonal.  Since
there are $4^\mu$ distinct vectors, they are a maximal set of
orthogonal vectors. We can therefore search among this set for the
optimal set of $N$ phase shift patterns to apply to our input horns.

As an example, consider the case $\mu=2$, that is, let the number of
time steps be $M=4^2=16$.  We will determine the maximum value of
$N$ that can be accommodated.  We proceed by assigning phase shift
sequences to the horns one at a time.   Without loss of generality,
we can assume that the first horn has no phase shift at all (since
any phase shift sequence can be subtracted from all inputs without
altering the solution):
\beq
\vec\phi_0=\langle
0,0\rangle=\vec\alpha_2^0\vec\alpha_1^0=\vec\alpha_0=
[0000000000000000]. \eeq
 Here $\vec\phi_j$ refers to the
$M$-dimensional vector
$(e^{i\phi_{j1}},e^{i\phi_{j2}},\ldots,e^{i\phi_{jM}})$. We can
accommodate two more inputs by choosing \beq \vec\phi_1=\langle
0,1\rangle=\vec\alpha_2^0\vec\alpha_1^1=
\vec\alpha_1=[0000111122223333], \eeq\beq \vec\phi_2=\langle
1,0\rangle=\vec\alpha_2^1\vec\alpha_1^0=
\vec\alpha_2=[0123012301230123].\eeq
For these three input horns, we
have six distinct baselines, with masks [equation (\ref{eq:Phi})]
\beq \vec\Phi_{01}=\langle 0,3\rangle,\quad \vec\Phi_{02}=\langle
3,0\rangle,\quad \vec\Phi_{12}=\langle 3,1\rangle, \eeq \beq
\vec\Phi_{10}=\langle 0,1\rangle,\quad \vec\Phi_{20}=\langle
1,0\rangle,\quad \vec\Phi_{21}=\langle 1,3\rangle. \eeq
These are
obtained by subtracting the values in angle brackets for the two
horns modulo 4.  for instance, $\vec\Phi_{12}=\langle 0,1\rangle-
\langle 1,0\rangle=\langle -1,1\rangle=\langle 3,1\rangle$. These
masks are all distinct, and hence mutually orthogonal, and
furthermore are all orthogonal to the vector $\langle 0,0\rangle$,
which is sensitive to the total power.

This construction shows that we can accommodate three horns with a
sequence of 16 time steps.  We next ask whether it is possible to
accommodate a fourth vector $\vec\phi_3$ in such a way that the new
masks $\vec\Phi_{03},\vec\Phi_{13},$ etc. are independent of the
ones we have already found. A search of the $16-7=9$ candidates
reveals an affirmative answer: $\vec\phi_3=\langle 3,3\rangle =
\vec\alpha_2^3\vec\alpha_1^3$ works.

The value $N=4$ is the maximum that can be achieved for the case of
$M=4^2$ time steps, as is clear from a counting argument: $N=5$
horns would require at least $M=N(N-1)+1=21$ steps.

\begin{table}
\begin{center}
\begin{tabular}{|c|c|c|}
\hline
$M$ & $N_{\rm max}$ (actual)& $N_{\rm max}$ (counting)\\
\hline
\hline
4 & 2& 2\\
\hline
$4^2=16$ & 4& 4\\
\hline
$4^3=64$ & 8& 8\\
\hline
$4^4=256$ & 15& 16\\
\hline
$4^5=1024$ & 24& 32\\
\hline
$4^6=4096$ & 40& 64\\
\hline
\end{tabular}
\end{center}
\caption{The number of horns $N$ that can be accommodated with a
given number of time steps $M$.  We assume $P=4$ distinct phase
shift values. The second column shows the maximum number that can be
accommodated, while the third column shows the number found by the
simple counting argument $N(N-1)+1\le M$.}
\label{table:nvsm}
\end{table}

Table \ref{table:nvsm} shows the maximum number of horns that can be
accommodated for various values of $M$.  These were found by
recursively searching the space of possible phase shifts in the
manner described above.  The last column shows the maximum value
that would be possible according to the simple counting argument
that the number of time steps must exceed the number of baselines.
We have repeated this analysis for the case $P=2$, where the phase
shifters are capable of only 0 and $180^\circ$ shifts, and found
very similar results for the relationship between $M$ and $N$.

As noted in the introduction,
extremely large values of $M$ are impractical.  This is
one reason that a bolometric interferometer with a large number of
horns should surely be designed with a high degree of symmetry, so
that there are many equivalent baselines that
can be read out coherently.  (The other reason
is the signal-to-noise advantage.)

In summary, this section has presented a procedure for selecting
$\vec\phi_j$ that yields fully orthogonal
masks. This means that the result of
applying the mask for a given baseline will only
be sensitive to the signal from the desired baseline, or equivalently that
the reconstruction of all visibilities is accomplished with minimal noise.

\section{Square Array}
\label{sec:Array}

We now consider the case where the input horns are arranged
in a square array with $N_{\rm side}$ horns on a side.
In this case, many different baselines (i.e.,
pairs of horns) sample the same visibility.  We wish to apply
identical phase shifts to such equivalent baselines, so that a single
mask reads out their sum.  Naturally, we also  require that
inequivalent baselines have orthogonal masks.  This
is the case considered in detail by Charlassier et al. (C08).
Our method parallels theirs in many respects but refines it in some
ways.

Following the notation of C08, we parameterize the
position of horns in the array in units of the minimum horn
separation as a vector $\vec d_j=(l_j,m_j)$.  Here $l_j,m_j$ are integers
running from 0 to $N_{\rm side}-1$, labeling the position
of the horn in along the $x$ and $y$ directions.
The index $j$ runs from 0 to $N_{\rm side}^2-1$ according to
$j=l_j+N_{\rm side}m_j$.
We can construct a set of phase shifts for all horns that satisfy
the desired criteria using the basis described in the previous
section with $\mu=2$.
We let the phase shift sequence for horn $j$ be
\beq
\vec\phi_{j}=\langle l_j,m_j\rangle.
\eeq


Below we have explicitly written out the modulations for each horn
in a $6 \times 6$ array.

\begin{displaymath}
\begin{array}{cccccc}
\langle 0,0 \rangle & \langle 1,0 \rangle & \langle 2,0 \rangle &
\langle 3,0 \rangle & \langle 4,0 \rangle & \langle 5,0 \rangle \\
\langle 0,1 \rangle & \langle 1,1 \rangle & \langle 2,1 \rangle &
\langle 3,1 \rangle & \langle 4,1 \rangle & \langle 5,1 \rangle \\
\langle 0,2 \rangle & \langle 1,2 \rangle & \langle 2,2 \rangle &
\langle 3,2 \rangle & \langle 4,2 \rangle & \langle 5,2 \rangle \\
\langle 0,3 \rangle & \langle 1,3 \rangle & \langle 2,3 \rangle &
\langle 3,3 \rangle & \langle 4,3 \rangle & \langle 5,3 \rangle \\
\langle 0,4 \rangle & \langle 1,4 \rangle & \langle 2,4 \rangle &
\langle 3,4 \rangle & \langle 4,4 \rangle & \langle 5,4 \rangle \\
\langle 0,5 \rangle & \langle 1,5 \rangle & \langle 2,5 \rangle &
\langle 3,5 \rangle & \langle 4,5 \rangle & \langle 5,5 \rangle \\
\end{array}
\end{displaymath}

In this case, the mask for the visibility corresponding to
any pair of horns is simply $\langle\Delta l,\Delta m\rangle$.
This means that all pairs with the same relative spacing get the same mask.
Furthermore, as long as the number of phase shift steps $P$ is
large enough, all inequivalent visibilities correspond
to orthogonal masks as desired.
The minimum value of $P$ is set by the fact that phases
are only defined modulo $P$.  Since $\Delta l,\Delta m$
can range from $-(N_{\rm side}-1)$ to $N_{\rm side}-1$
we need at least $P=2(N_{\rm side}-1)+1$ distinct phase shifts.
(As we will see in the next section, it may be desirable for $P$
to be a multiple of 3, in which case we simply round
up to the nearest such value.)
If $P$ is smaller than this, then distinct visibilities will
be mapped onto the same phase shift sequence.
For the above case, for example, we require $P\ge 11$.
If we tried a smaller value, say $P=10$, then
the visibility corresponding to horns (0,0) and (5,1), for example, would
get the same phase shift sequence as (5,0) and (0,1), namely
$\langle 5,1\rangle=\langle -5,1\rangle$.


It is instructive to compare this scheme with the very similar one
of C08.  In both methods, the phase shift sequence
for horn $(l,m)$ is expressed in the form $l\vec h+m\vec v$ for two
basis shift patterns $\vec h,\vec v$.  In order to achieve the desired
orthogonality properties, Charlassier et al. choose $\vec h,\vec v$ to
be independent random vectors of phase shifts.  The randomness
ensures approximate
orthogonality, up to errors of order $M^{-1/2}$, where $M$ is the
length of the phase shift sequence.  In contrast, we choose
$\vec h=\langle 1,0\rangle$ and $\vec v=\langle 0,1\rangle$.  This
results in strict orthogonality, as opposed to approximate orthogonality.

\begin{figure}
\centerline{\includegraphics[width=3in]{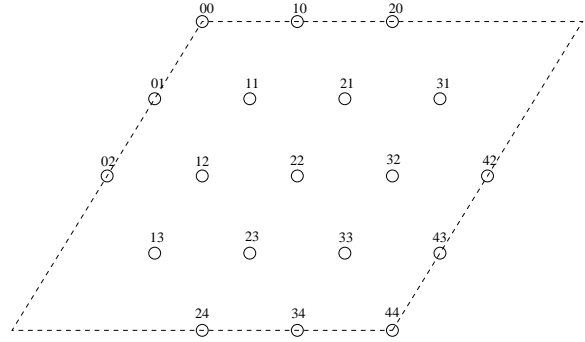}} \caption{A
hexagonal array of 19 horns can be seen as a subset of a $5\times 5$
parallelogram-shaped array.  The phase shifting scheme described in
Sect.~\ref{sec:Array} with $N_{\rm side}=5$ can be applied to this
array.} \label{fig:hex}
\end{figure}

The number $P$ of distinct phase shift values required is essentially
the same
in the two methods.  As in our method,
C08 found that $P\simeq 2N_{\rm side}$ was required
in order to produce orthogonal phase shifts using random basis vectors.

Using either method, the length $M$ of the modulation sequences
is greatly reduced compared to the case of inequivalent
baselines.  The number of required phase shifts
is $M=P^2=(2N_{\rm side}-1)^2$ when redundant baselines are tagged
equivalently.  If we instead used the methods of the previous section,
we would require $M>N_{\rm side}^2(N_{\rm side}^2-1)+1\approx N_{\rm side}^4$.
For the $6\times 6$ array denoted above, this is the difference
between a 121-step sequence and a 1261-step sequence.
Even more important is the signal-to-noise benefit of coadding
equivalent baselines.


Although we have described this procedure as applying to a square array,
it is in fact more general.  It applies whenever the horn positions
can be expressed as integer multiples of any two basis vectors,
even if the two are not orthogonal, or in other words, to any
parallelogram-shaped array.  Furthermore, it can be applied to
any subset of such a parallelogram-shaped array, since we can simply
ignore the parts of the array with no horns in them.
In particular, this means that the method can be applied to
a hexagonal close-packed array of horns, as shown in Fig.~\ref{fig:hex}.


\section{Two Polarizations}
\label{sec:Pol}
Thus far, for simplicity we have been considering only one
polarization state of the incoming radiation field.  We now imagine
that two orthogonal linear polarizations $(x,y)$ are measured at each horn.
In this case, we can in principle recover visibilities for all
four Stokes parameters $I,Q,U,V$.  To be specific, if visibilities
$V_{xx},V_{xy},V_{yx},V_{yy}$ are measured for a particular baseline,
then
\begin{eqnarray}
V_I&=&V_{xx}+V_{yy},\label{eq:vi}\\
V_Q&=&V_{xx}-V_{yy},\label{eq:vq}\\
V_U&=&V_{xy}+V_{yx},\label{eq:vu}\\
V_V&=&-i(V_{xy}-V_{yx}).\label{eq:vv}
\end{eqnarray}
However, as C08 have pointed out,
it is impossible to recover all visibilities while taking full advantage
of the noise reduction resulting from coadding redundant baselines.
C08 describe two schemes for recovering some of the Stokes parameters
with full accuracy, one of which (mode 2 of C08) involves measuring
the visibilities for Stokes $I,U,V$ but not $Q$.
Stokes $Q$ can then be measured by rotating the instrument $45^\circ$.
In this section, we show how to implement this mode of operation
using our phase shifting scheme.

Aside from the phase shifting scheme,
there is another reason for adopting an observing scheme in which
Stokes $Q$ is measured only by rotating the instrument.
As eq.~(\ref{eq:vq}) indicates, the visibility for Stokes
$Q$ is obtained by subtracting two measured visibilities, each
of which
contains a contribution proportional to the much larger Stokes parameter $I$.
As a result, this visibility is likely to be subject to much larger errors than
the other linear polarization (Stokes $U$).

As in the previous section, we assume an $N_{\rm side}\times N_{\rm side}$
array of horns,
but now we introduce an orthomode transducer for each horn, doubling
the number of signals to be interfered. We can represent each of these
$2N_{\rm side}^2$ signals with a triple of labels $(l_j,m_j,n_j)$ where
$(l_j,m_j)$ label the position of the horn as in the previous section,
and $n_j=0,1$ labels the polarization state.
In the previous section, we identified the horn $(l_j,m_j)$ with
a phase shift sequence $\langle l_j,m_j\rangle$.
In the present case,
we can define phase shift sequences similarly in terms of the triple
$\langle l_j,m_j,n_j\rangle$, each of which represents a
a sequence of
$3P^2$ time steps, where $P\ge 2N_{\rm side}-1$  as in the previous
section. For one of the two polarization states, we the phase shift
sequences are simply three repetitions of the sequences we found
previously:
\begin{equation}
\langle l,m,0\rangle =\left(\langle l,m\rangle,\langle l,m\rangle,
\langle l,m\rangle\right).
\end{equation}
For the other polarization state, we apply a slow three-phase
modulation to this sequence: the first $P^2$ steps are unchanged,
the next $P^2$ steps are multiplied by $e^{2\pi i/3}$, and the final
block is multiplied by $e^{4\pi i/3}$:
\begin{equation}
\langle l,m,1\rangle =\left(\langle l,m\rangle,e^{2\pi i/3}\langle
l,m\rangle, e^{4\pi i/3}\langle l,m\rangle\right).
\end{equation}
Note that this scheme is most natural to apply when $P$ is a
multiple of 3 so that for every phase state $p$ there is another
whose phase is $p+2\pi/3$. Otherwise, the set of phase shifts
involved in the sequences $\langle l,m,1\rangle$ will be larger than
that involved in $\langle l,m,0\rangle$.  In implementing this
scheme, one would surely round $P$ up to the nearest multiple of 3.

It is straightforward to check that
all equivalent baselines have identical phase shift sequences as
desired. All pairs that interfere $x$ and $y$ polarization have
independent, orthogonal phase shift sequences, allowing optimal
reconstruction of Stokes $U,V$ visibilities
[eqs.~(\ref{eq:vu}),(\ref{eq:vv})]. Those that interfere $x$ and $x$
have identical sequences to those that interfere $y$ and $y$.
Applying these phase shift masks therefore allows recovery of the
sum of these visiblities, which is $V_I$.


As in the previous section, this method is similar to that of
C08, except that our method imposes strict orthogonality
on distinct baselines, as opposed to relying on the approximate
statistical orthogonality that results from choosing random
phase shift sequences.


As an example, consider a square array with $N_{\rm side}=8$. The
number of different phase shift values must satisfy $P\ge 2N_{\rm
side}-1 =15$. The length of the phase shift sequences is
$M=3P^2=675$. The shortest sequence of phase shifts we could
possibly hope for would have $M$ equal to the number of unknowns we
are trying to solve for.  In this arrangement, there are 112
inequivalent baselines, each of which has three complex visibilities
that are measured, and in addition the total power in $I,Q,U$ are
measured, resulting in a total of $6\times 112+3=675$ unknowns. Our
phase shift sequence is therefore as short as possible. For
comparison, according to Fig.~4 of C08, the optimal noise levels in
the C08 scheme are obtained only when the phase shift sequence is
$\sim 3$ times the minimum length. In these comparisons we only
consider the length of mode 2 in C08. It should be noted that when
$2N_{\rm side}-1$ is divisible by 3 that we recover visibilities
with maximum efficiency. When this is not the case the ratio of our
length to the minimum approaches 1 for large $N_{\rm side}$. For
arrays of reasonable size ($8\times 8$ or larger) the maximum ratio
is 1.22 and occurs for a $10\times10$ array.

\section{Conclusions}
\label{sec:Conclusions} Optimal recovery of visibilities in a
bolometric interferometer depends on the proper choice of phase
shift sequences. We have laid out a method for finding such
sequences that lead to fully orthogonal masks for all visibilities
and introduced a compact notation for describing such phase shift
sequences.

In the case of an array with a regular lattice structure, equivalent
baselines can be read out simultaneously, reducing the length of the
required phase shift sequence and improving the signal-to-noise
ratio.  This method refines that of C08.  The method applies to
arrays that are based on replication of any parallelogram-shaped
fundamental cell, including for example hexagonal arrays.

For the case of rectangular arrays, the method described herein is
very similar to that of C08, although our method imposes strict
orthogonality on the masks for distinct baselines, rather than
relying on approximate orthogonality resulting from random
sequences.  As a result, our method leads to optimal recovery of
visibilities for Stokes' $I,U,V$, with shorter time sequences than
that of C08.

The ability to shorten the sequence of phase shifts is likely to be
important in instrument design, because it reduces the degree to
which $1/f$ noise must be controlled.  For example, suppose that we
can shift phase states at a rate of one state per 10 ms (either
because of the design of the phase shifters or the bolometer time
constants).  As we saw in the previous section, an $8\times 8$ array
requires $\sim 1000$ phase shifts, which would take 10 seconds.  We
therefore require the $1/f$ noise knee to be below $\sim 0.1$ Hz.
An alternative scheme involving a longer phase shift sequence would
require correspondingly tighter control of the $1/f$ knee.

\begin{acknowledgements}
EFB is supported by National Science Foundation Award AST-0507395.
During the development of this paper POH was supproted by a NASA
awards NAG5-12758 and NNX07AG82G and a grant from the Wisconsin
Space Grant Consortium. We thank Peter Timbie, Andrei Korotkov, Greg
Tucker, and the other members of the MBI collaboration, as well as
Roman Charlassier, Jean-Christophe Hamilton, and Jean Kaplan, for
valuable discussions.
\end{acknowledgements}

\bibliographystyle{hapj}
\bibliography{modpaper}

\appendix
\section{Proof of minimum-noise condition}
\label{sec:appendix}

In this section we provide a proof of the assertion that orthogonal
phase shift patterns minimize the noise in the recovered
visibilities.

Assume that the visibilities are arranged in an $N_{\rm
vis}$-dimensional
 vector $\vec V$, and
the observed signals are arranged in an $M$-dimensional vector $\vec
S$. The two are related by an $M\times N_{\rm vis}$ matrix ${\cal A}$:
\beq
\vec S={\cal A}\vec V
\eeq
All entries of ${\cal A}$ are complex numbers with
absolute value 1.  We assume that $M\ge N_{\rm vis}$ and that the
matrix ${\cal A}$ has maximal rank, so that it is possible to solve for the
unknown visibilities.

Assuming that the signals are contaminated with white noise with
variance $\sigma^2$, the optimal reconstruction of the visibilities
is the least-squares vector
\beq
\hat{\vec V}=({\cal A}^\dag
{\cal A})^{-1}{\cal A}^\dag\vec S.
\eeq
The noise covariance matrix for $\hat{\vec
V}$ is
\beq
{\cal N}=\sigma^2({\cal A}^\dag {\cal A})^{-1}.
\eeq
The noise in the
$j$th recovered visibility has variance ${\cal N}_{jj}$. We wish to
show that this noise is minimized when the matrix ${\cal A}$ has orthogonal
columns.

The diagonal elements of the inverse noise matrix are
\beq
({\cal N}^{-1})_{jj}=\sigma^{-2}({\cal A}^\dag {\cal A})_{jj}=
\sigma^{-2}\sum_{m=1}^M
A^*_{mj}A_{mj}= {M\over\sigma^2}.
\eeq
We can therefore write
\beq
{\cal N}^{-1}={M\over\sigma^2}({\cal I}+{\cal D}),
\eeq
where ${\cal I}$ is the
identity matrix and ${\cal D}$ is a hermitian matrix with zeroes along
the diagonal.

The noise covariance matrix is
\beq
{\cal N}={\sigma^2\over
M}({\cal I}-{\cal D}+{\cal D}^2-{\cal D}^3+\ldots) ={\sigma^2\over
M}[{\cal I}-{\cal D}+{\cal D}(1-{\cal D}+{\cal D}^2-\ldots){\cal D}]
={\sigma^2\over
M}({\cal I}-{\cal D}+{\cal D}{\cal N}{\cal D}).
\eeq
Since ${\cal D}$ has no
diagonal elements, an arbitrary diagonal element of the noise
covariance matrix is
\beq
{\cal N}_{jj}={\sigma^2\over
M}[1+({\cal D}{\cal N}{\cal D})_{jj}] ={\sigma^2\over M}[1+\vec
v^\dag{\cal N}\vec v],
\eeq
where $\vec v_k=D_{kj}$.  Since
${\cal N}$ is a positive definite matrix, we conclude that \beq
{\cal N}_{jj}\ge {\sigma^2\over M}. \eeq That is, the minimum noise
variance achievable on any one visibility is $\sigma^2/M$.  This
value is achieved when the matrix ${\cal A}$ is column orthogonal, since in
this case ${\cal A}^\dag {\cal A}=(M/\sigma^2){\cal I}$ and
${\cal N}=( \sigma^2/M){\cal I}$.

\end{document}